\pdfoutput=1
\documentclass[aps,pre,preprint,groupedaddress,nofootinbib]{revtex4-1}
\usepackage[]{natbib}
\usepackage{amsmath, amssymb, amsthm, url}
\usepackage{graphicx}
\usepackage{xcolor}  % Allow for color text
\usepackage[normalem]{ulem} % Allow for strikeout text (\sout)
\usepackage{dcolumn}
\newcolumntype{d}[1]{D{.}{.}{#1}}
\begin{document}

%\preprint{}

\title{The tipping point: a mathematical model for the profit-driven abandonment of restaurant tipping}
%SHORT TITLE \title{Handicap principle explains dimorphic ornaments}



\author{Sara M.~Clifton}
\email[E-mail me at: ]{smc567@illinois.edu}
\affiliation{Department of Mathematics, University of Illinois at Urbana-Champaign, Urbana, IL 61801, USA}

\author{Eileen Herbers}
\affiliation{Department of Engineering Sciences and Applied Mathematics, Northwestern University, Evanston, IL 60208, USA}

\author{Jack Chen}
\affiliation{Department of Engineering Sciences and Applied Mathematics, Northwestern University, Evanston, IL 60208, USA}

\author{Daniel M.~Abrams}
\affiliation{Department of Engineering Sciences and Applied Mathematics, Northwestern University, Evanston, IL 60208, USA}
\affiliation{Northwestern Institute for Complex Systems, Northwestern University, Evanston, IL 60208, USA}
\affiliation{Department of Physics and Astronomy, Northwestern University, Evanston, IL 60208, USA}

%\keywords{ornament | evolution | natural selection | sexual selection | competition | speciation | handicap principle | mathematical model}

\date{\today}

\begin{abstract}
The custom of voluntarily tipping for services rendered has gone in and out of fashion in America since its introduction in the 19th century. Restaurant owners that ban tipping in their establishments often claim that social justice drives their decisions, but we show that rational profit-maximization may also justify the decisions. Here, we propose a conceptual model of restaurant competition for staff and customers, and we show that there exists a critical conventional tip rate at which restaurant owners should eliminate tipping to maximize profit. Because the conventional tip rate has been increasing steadily for the last several decades, our model suggests that restaurant owners may abandon tipping en masse when that critical tip rate is reached.
\end{abstract}

\maketitle

%\linenumbers

\section{Introduction}
Tipping for restaurant service has gone in and out of fashion in America since its introduction from Europe in the 19th century \cite{azar2004history}. Tipping has always been a controversial social convention, for both scholars and the public. The practice has been consistently tied to the worst of human nature: racism, sexism, and classism \cite{lynn2008consumer,rind1996effect,lynn2000predictors,margalioth2006case}. At many points in time, tipping has been considered downright anti-democratic \cite{segrave1998tipping}. Yet the practice persists because the vast majority of Americans prefer to choose how much gratuity they leave after a meal \cite{lynn2015effects,azar2010tipping}.

Economists have traditionally struggled to explain the practice of tipping in terms of rational costs and benefits because a rational economic agent would not incur a monetary cost that provides no present or future benefit \cite{azar2004sustains}. Sociologists and psychologists have appealed to negative feelings of guilt, embarrassment, or anxiety to explain why people conform to the social convention of tipping \cite{elster1989social,lynn2009individual,lynn1993consumer}, while others have appealed to altruistic feelings of generosity and empathy \cite{greenberg2014complementarity,strohmetz2002sweetening,fong2005socio}.

While much scholarly attention has been paid to the consumers who tip and employees who receive tips (e.g. \cite{lynn1990restaurant,gibson1997economics,rind1995effect,lynn1993effect}), relatively little has been paid to restaurant owners who employ tipped staff. Theoretical models of tipping often see restaurant owners as inefficient judges of service quality \cite{conlin2003norm}. The natural conclusion to this rationale is that restaurant owners should allow tips in their restaurants in order to efficiently evaluate the quality of their wait staff \cite{lynn2001restaurant}. This conclusion breaks down when we consider the real-world factors that influence restaurant owners' decisions, such as remaining profitable while also adhering to minimum wage restrictions, retaining talented staff, and meeting customer expectations for food and service quality \cite{enz2004issues}. By accounting for these factors and appealing only to rational profit-maximization motivations, we show that restaurants owners should play a more active role in determining tip rates in their restaurants.  

%Recently, prominent restaurant owners have abolished tipping in their restaurants, calling the practice fundamentally unfair and igniting a fresh debate about gratuity across the country \cite{eater:tip2}. While Danny Meyer and other restaurant owners cite social justice as the primary reason to ban tipping in their establishments, may there exist a more selfish justification for the decision? 

\begin{figure}[htb]
  \begin{center}
    \includegraphics[width=0.55\textwidth]{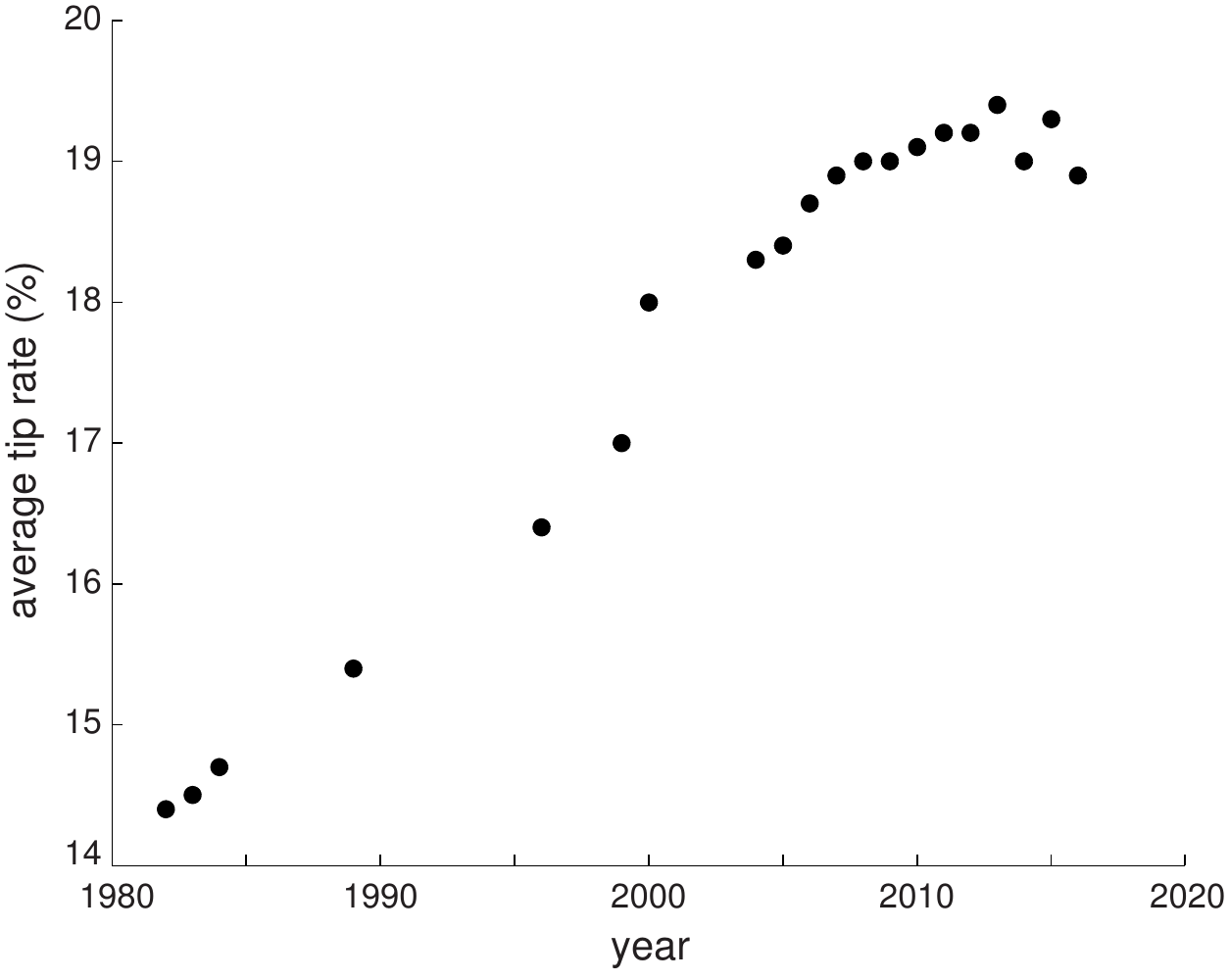} \end{center}
  \caption[Average reported tip rate in American restaurants over time.]{Average reported tip rate in American restaurants over time, according to NPD Group (1982-84) and Zagat annual surveys (1989-present) \cite{Margalioth:2010}. Note that both surveys are aimed at diners who patronize full-service midscale and upscale restaurants.}
  \label{fig1}
\end{figure}

As the conventional tip rate gradually increases in the US (see Fig \ref{fig1}), waiters' take home pay steadily increases, while back-of-house employees' pay remains stagnant \cite{nyt:whytip}. Despite the low federal minimum wage for tipped workers (\$2.13 as of 2017 \cite{minwage}), waiters consistently earn more than cooks \cite{lynn2017should}. As the wage disparity increases, talented cooks may defect to restaurants where profits are shared more equitably among staff, and talented waiters may defect to restaurants with higher tips. A rational restaurant owner interested only in maximizing profit might take control of the tip rate in his/her restaurant in order to retain the most talented front-of-house (tipped) and back-of-house (untipped) staff. We show in a conceptual model of two competing restaurants that a critical tipping rate exists at which a rational restaurant owner will abandon tipping to maximize profit. 

%%%%%%% TIPPING FIGURE %%%%%%%
%%%%%%% TIPPING FIGURE %%%%%%%

\section{Methods}
% OUR MODEL
\subsection{Model derivation}
As a simple conceptual model, consider two restaurants competing for diners, waiters, and cooks. For notational ease, we will focus on one restaurant (\textit{our restaurant}, Restaurant 1). We assume all diners, waiters, and cooks intend to eat or work, respectively, at either our restaurant or the competing restaurant. Following a dynamical systems approach by Abrams et al.~for modeling generic social group competition \cite{abrams2011dynamics}, people will transition between restaurants at a rate proportional to the relative utility of being at each restaurant (either as a customer or an employee). This general model assumes that the popularity of a social group also influences the transition rates, but for our purpose of modeling purely rational behavior, we focus only on utility as a driving force. Our simplification reduces the model by Abrams et al.~to
\begin{align} \label{abrams}
\frac{\mathrm{d} x}{\mathrm{d} t} = y \, P_{y x} (u_x) - x \, P_{x y} (u_x),
\end{align}
where $x$ is the fraction of people in group $X$, $P_{yx}$ is the probability per unit time of transitioning from group $Y$ to group $X$, $u_x$ is the utility of group $X$, and transition rates are symmetric under exchange of $x$ and $y$. In our case, $X$ is our restaurant and $Y$ is the competing restaurant. Alternatively, the ``competing restaurant" could be viewed as a reservoir of all other dining options (including home), but that extension is left for future work.

\subsubsection{Dynamics of cooks}
Suppose our system has a number $N_C$ of cooks who must choose between our restaurant and the competing restaurant. Many factors influence a cook's decision to work at a particular restaurant, but we model cooks primarily concerned with pay. Because cooks do not receive tips, cooks choose where to work based only on the relative wage at the two restaurants. Assuming for simplicity that the transition probabilities are linear in relative base cook pay, the change in the number of cooks $\tilde{C}$ at our restaurant is
\begin{align}
\tau_C \frac{\mathrm{d} \tilde{C}}{\mathrm{d} t} =  \underbrace{(N_C - \tilde{C}) \, \frac{b_{C1}}{b_{C1} + b_{C2}}}_{\substack{\text{switch from} \\ \text{competitor to us}}} - \underbrace{\tilde{C} \,  \frac{b_{C2}}{b_{C1} + b_{C2}}}_{\substack{\text{switch from} \\ \text{us to competitor}}}, \label{eq:cooktilde}
\end{align}
where $\tau_C$ sets the transition time scale for cooks, and $b_{C1}$ and $b_{C2}$ are the hourly base cook pay at our restaurant and the other restaurant, respectively. For example, if both restaurants offer the same base pay ($b_{C1}=b_{C2}$), then eventually half the cooks will be at our restaurant, and the other half will go to the competitor.  Tilde notation will be removed later when the model is normalized. 

\subsubsection{Dynamics of waiters}
Like the cooks, the waiters in our system are primarily concerned with pay. Because waiters receive both hourly wages and gratuity, the transition rate between restaurants depends on the relative hourly take home (total) pay at the two restaurants. The hourly gratuity at our restaurant is
\begin{align} \label{gratuity}
g_1 = \frac{m_1 \tilde{D} T_1}{\tilde{W}},
\end{align}
where $m_1$ is the hourly menu price, $\tilde{D}$ is the number of diners, $T_1$ is the tip rate, and $\tilde{W}$ is the number of waiters that must split the total tips. The gratuity $g_2$ at the competing restaurant is similarly defined.

The hourly take home pay at our restaurant is then $b_{W1} + g_1$, where $b_{W1}$ is the hourly base waiter pay at our restaurant. The change in the number of waiters $\tilde{W}$ at our restaurant is then
\begin{align}
\tau_W \frac{\mathrm{d} \tilde{W}}{\mathrm{d} t} = \underbrace{(N_W - \tilde{W}) \, \frac{b_{W1} + g_1}{b_{W1} +g_1+b_{W2}+g_2}}_{\substack{\text{switch from} \\ \text{competitor to us}}} - \underbrace{\tilde{W} \,  \frac{b_{W2}+g_2}{b_{W1}+g_1+b_{W2}+g_2}}_{\substack{\text{switch from} \\ \text{us to competitor}}}, \label{eq:waitertilde}
\end{align}
where $\tau_W$ sets the transition time scale for waiters, and $N_W$ is the number of waiters in the system.

\subsubsection{Dynamics of diners}
Assuming all diners intend to eat at a restaurant, they must chose between our restaurant and our competitor. Many factors influence a person's decision to eat at a particular restaurant, but we will focus on food and service quality versus menu cost. There are also many ways to measure food and service quality \cite{sulek2004relative,namkung2008highly,gagic2013vital}, but we will use the number of cooks and waiters who choose to work at our restaurant as a basic proxy. For instance, if our restaurant attracts more waiters, then diners will receive more personal attention and perceived service quality will increase. Suppose for simplicity that the quality $q_1$ of the meal and service at our restaurant is a linear combination of the number of cooks $\tilde{C}$ and waiters $\tilde{W}$ working at our restaurant:
\begin{align} \label{quality}
q_1 = \alpha_W \tilde{W} + \alpha_C \tilde{C},
\end{align}
where $\alpha_W$ and $\alpha_C$ are the weights placed on service and food, respectively, when evaluating our restaurant.  The quality $q_2$ of the competing restaurant is defined similarly. Note that this proxy for restaurant quality will only be an acceptable approximation if both restaurants are not grossly over- or under-staffed. We explore two alternative formulations for restaurant quality in the supplementary online material, and we find similar qualitative results. 

We define the value $v_1$ of our restaurant as the quality $q_1$ over the menu cost (including tips): 
\begin{align} \label{value}
v_1 = \frac{\alpha_W \tilde{W} + \alpha_C \tilde{C}}{m_1(1+T_1)},
\end{align}
where $m_1$ is the hourly menu cost and $T_1$ is the tip rate at our restaurant. The value $v_2$ of the other restaurant is defined similarly.

A rational diner chooses a restaurant based on the perceived relative value of each restaurant. The change in the number of diners $\tilde{D}$ at our restaurant is then
\begin{align}
\tau_D \frac{\mathrm{d} \tilde{D}}{\mathrm{d} t} = \underbrace{(N_D - \tilde{D}) \, \frac{v_1}{v_1+v_2}}_{\substack{\text{switch from} \\ \text{competitor to us}}} - \underbrace{\tilde{D} \,  \frac{v_2}{v_1+v_2}}_{\substack{\text{switch from} \\ \text{us to competitor}}}, \label{eq:dinertilde}
\end{align}
where $\tau_D$ sets the transition time scale for diners, and $N_D$ is the number of diners in the system. 

\subsubsection{Profitability}
Given the flow of employees and customers to and from our restaurant, a rational restaurant owner will maximize hourly profit
\begin{align} \label{profittilde}
\tilde{P} = \underbrace{m_1 \, \tilde{D}}_{\text{revenue}} - \underbrace{b_{W1} \, \tilde{W}}_{\text{waiter pay}} - \underbrace{b_{C1} \, \tilde{C}}_{\text{cook pay}}.
\end{align}
We ignore fixed costs because we are only concerned with maximizing profitability, not absolute profits.

\subsection{Normalized model}
We now normalize and nondimensionalize the system \eqref{eq:cooktilde}-\eqref{eq:dinertilde} to reduce the number of parameters. We make the following substitutions
\begin{align}
D &= \frac{\tilde{D}}{N_D}, \,\,\, W = \frac{\tilde{W}}{N_W}, \,\,\, C = \frac{\tilde{C}}{N_C} \\
r &= \frac{\alpha_C}{\alpha_W}, \,\,\, r_{DW} = \frac{N_D}{N_W}, \,\,\, r_{CW} = \frac{N_C}{N_W},
\end{align}
so that $D$, $W$, and $C$ are the fraction diners, waiters and cooks at our restaurant, $r$ is the ratio of food to service importance for customers, and $r_{DW}$ and $r_{CW}$ are the ratios of diners and cooks to waiters, respectively. We also rescale time such that $\tau_C=\tau_W=\tau_D = 1$. Naturally, the transition rates may vary for diners, waiters, and cooks; customers may switch dining locations more rapidly than employees switch jobs. However, we are only interested in equilibrium states, so we ignore this detail.

Then the fraction of diners at our restaurant follows the dynamics
\begin{align}
\frac{\mathrm{d} D}{\mathrm{d} t} =&  \, (1 - D) \, \frac{v_1}{v_1 + v_2} - D \,  \frac{v_2}{v_1 + v_2}  \label{eq:diner} \\
v_1 = & \, \frac{W+ r \, r_{CW} \, C}{m_1(1+T_1)}, \,\,\,\,\, v_2 = \, \frac{(1-W) + r \, r_{CW} \, (1-C)}{m_2(1+T_2)}. \label{eq:value}
\end{align}
The fraction of waiters at our restaurant follows the dynamics
\begin{align}
\frac{\mathrm{d} W}{\mathrm{d} t} =& \, (1 - W) \, \frac{b_{W1} + g_1}{b_{W1} + g_1 + b_{W2} + g_2} - W \,  \frac{b_{W2} + g_2}{b_{W1} + g_1 + b_{W2} + g_2}  \label{eq:waiter} \\
g_1 = &  \, \frac{m_1 r_{DW} D T_1}{W}, \,\,\,\,\, g_2 = \, \frac{m_2 r_{DW} (1-D) T_2}{W}. 
\end{align}
Finally, the fraction of cooks at our restaurant follows the dynamics
\begin{align}
\frac{\mathrm{d} C}{\mathrm{d} t} =  \, (1 - C) \, \frac{b_{C1}}{b_{C1} + b_{C2}} - C \,  \frac{b_{C2}}{b_{C1} + b_{C2}}. \label{eq:cook}
\end{align}
All variables and parameters are described in Table \ref{tab:paramtip}.

%%%%%%%%% Parameter table %%%%%%%%
	\begin{table}[!ht]
	\begin{center}
	\begin{tabular}{| c  p{9.4cm} p{1.3cm} p{1.75cm}  p{1.75cm} |}  \hline 
	{\bf Variable} & {\bf Meaning} & {\bf Units} & {\bf Range} & {\bf Baseline} \\  \hline 
	$D$ & fraction of diners at our restaurant & -- & [0, 1] & --  \\ 
	$W$ & fraction of waiters at our restaurant & -- & [0, 1] & --  \\  
	$C$ & fraction of cooks at our restaurant & -- & [0, 1]  & -- \\ 
	$t$ & time (dimensionless) & -- & [0, $\infty$) & -- \\
	$r$ & relative importance of food quality versus service quality, typically a value exceeding one & -- & [4, 20]  & 12$\S$ \\ 
	$r_{CW}$ & ratio of total cooks to waiters in the system & -- & [0.5, 2] & 1  \\ 
	$r_{DW}$ & ratio of total diners to waiters in the system & -- & [1, 20]  & 12 \\ 
	$m_1$ & average menu cost per hour at our restaurant & \$/hr & [5, 20] & 10  \\ 
	$b_{W1}$ & waiters' base pay per hour at our restaurant & \$/hr & [2.13*, 25] & 5.00\dag   \\ 
	$b_{C1}$ & cooks' base pay per hour at our restaurant & \$/hr & [7.25*, 25] & 10.40\dag \\ 
	$T_1$ & average tip rate at our restaurant, determined by either social convention or mandated by restaurant owner & -- & [0.01, 0.5] & 0.19\ddag \\ 	
	$v_1$ & meal value perceived by customers at our restaurant & 1/\$ & -- & --  \\ 
	$g_1$ & gratuity per hour at our restaurant & \$/hr & -- & -- \\ 
	\hline
	\end{tabular} \end{center}
	\caption{Description of model variables and parameters for our restaurant, Restaurant 1. The competing restaurant (Restaurant 2) has similarly defined parameter values subscripted with 2. We present a range of plausible values for each parameter and a baseline value for midscale and upscale restaurants like those reviewed by Zagat.  ($\S$ crude estimate based on customer surveys \cite{sulek2004relative}; *federal minimum wage as of 2017 \cite{minwage}; \dag average waiter and cook pay as of 2015 \cite{wagecomp}; \ddag average self-reported tip rate as of 2016 \cite{Margalioth:2010}; other baseline values are guesses based on author experience).}
	\label{tab:paramtip}
	\end{table}
%%%%%%%%% Parameter table %%%%%%%%

With this change of variables, hourly profit $\tilde{P}$ becomes the hourly profit per waiter in the system
\begin{align} \label{profit}
P = \, m_1 \, r_{DW} \, D - b_{W1} \, W - b_{C1} \, r_{CW} \, C.
\end{align}

\section{Results}

\subsection{Numerical exploration}
Numerical integration suggests that one stable steady state solution exists for each set of parameters regardless of the initial condition, so long as the initial condition is physically meaningful. Because cooks only switch restaurants in response to base pay (constant parameter), the distribution of cooks equilibrates first. Diners and waiters respond to everyone else in the system, so the distribution of diners and waiters equilibrates later. 

\begin{figure}[htb]
  \begin{center}
    \includegraphics[width=0.72\textwidth]{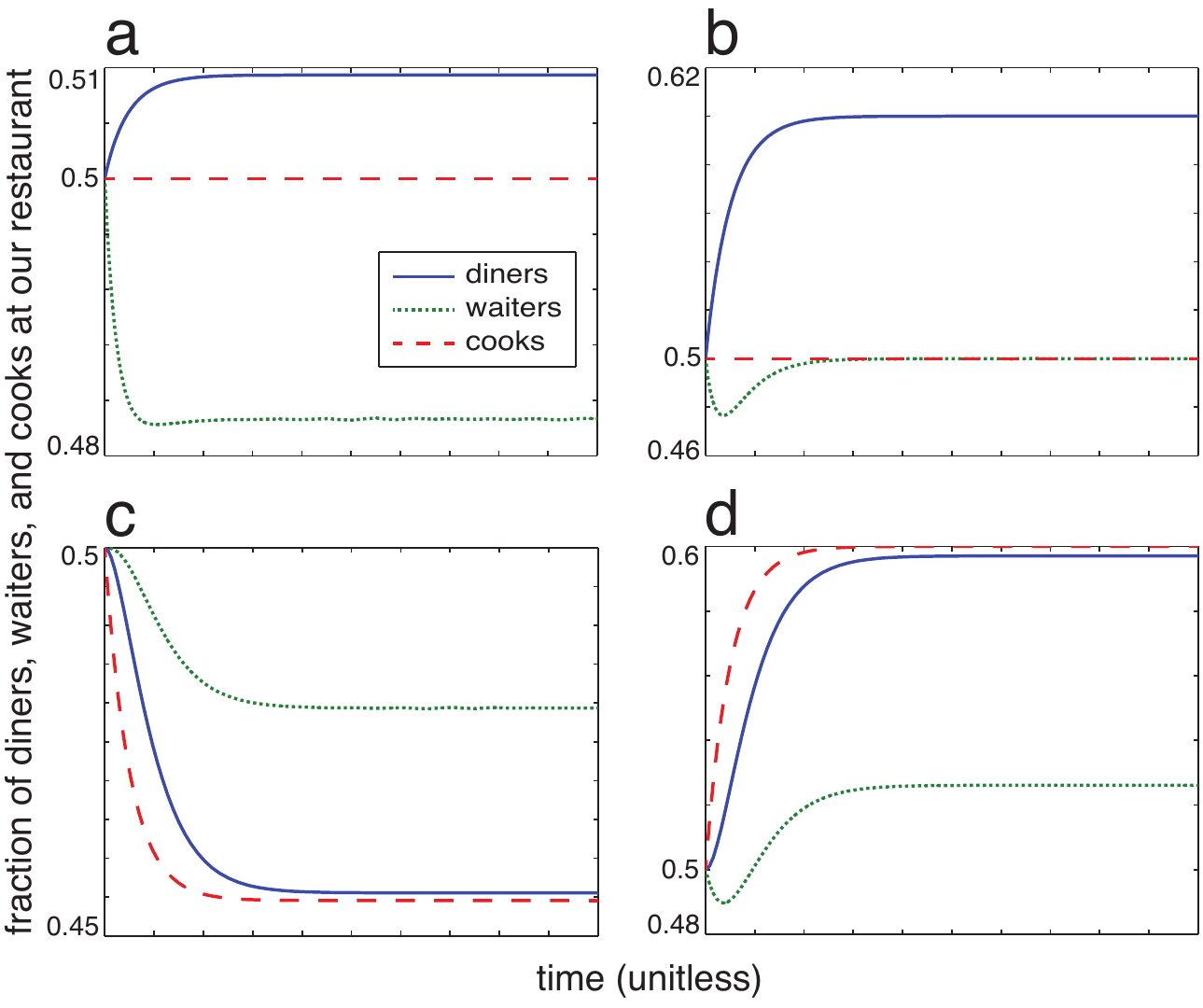} \end{center}
  \caption[Numerical simulation of system.]{(Color online) Numerical simulation of system \eqref{eq:diner}-\eqref{eq:cook} with two nearly identical competing restaurants. \textbf{(a)} If the tip rate at our restaurant is lower than the competitor, then waiters will leave, but diners will prefer our restaurant ($T_2=0.25$). \textbf{(b)} If the menu price at our restaurant is lower than the competitor, then diners will flock to our restaurant, and waiters will temporarily leave. However, after our restaurant has a large share of diners, waiters will return ($m_2=15$). \textbf{(c)} If we pay our cooks less, then cooks followed by diners followed by waiters will leave ($b_{c2}=12$). \textbf{(d)} If we pay our cooks more but pay our waiters less to compensate, cooks will flock to our restaurant followed by diners; waiters will temporarily leave because they are paid lower wages, but eventually they will come back as diners flood our restaurant ($b_{w2}=10, b_{c1}=15$). Unless otherwise noted, $m_1=m_2=10, T_1=T_2=0.2, b_{w1}=b_{w2}=5, b_{c1}=b_{c2} =10, r=12$.}
  \label{fig2}
\end{figure}

For otherwise identical restaurants, small changes in restaurant policy (like staff pay, menu prices, and tip rates) will have an effect on the entire restaurant ecosystem. Lowering the tip rate at our restaurant will cause waiters will leave our restaurant because they get paid less, but diners will prefer our restaurant because they pay less (see Fig \ref{fig2}a). 

If the menu price at our restaurant is lower than the competitor, then diners will flock to our restaurant because they pay less, and waiters will temporarily leave our restaurant because lower menu prices lead to lower tips. However, after our restaurant has a large share of diners, waiters return because the density of diners balances the lower menu prices (see Fig \ref{fig2}b). 

If we pay our cooks less than our competitor, then cooks will leave our restaurant because they get paid less; as food quality decreases, diners will leave our restaurant, and then waiters will leave our restaurant as their hourly tips decrease (see Fig \ref{fig2}c). 

As a final example, if we pay our cooks more but pay our waiters less to compensate, cooks will flock to our restaurant followed by diners; waiters will temporarily leave because they are paid lower wages, but eventually they will come back as diners flood our restaurant (see Fig \ref{fig2}d).

%%%%%%% SIMULATION FIGURE %%%%%%%
%%%%%%% SIMULATION FIGURE %%%%%%%

\subsection{Equilibrium stability analysis}
Fixed point analysis shows that four steady states exist. Only one fixed point is meaningful (i.e., $D^*,W^*,C^* \in [0,1]$). The steady state for cooks is $C^* =  b_{C1}/(b_{C1}+b_{C2})$. The steady states for waiters and diners have closed forms but are too long to include. For all reasonable parameter values (listed in Table \ref{tab:paramtip}), the eigenvalues of the Jacobian evaluated at the fixed point are real and negative. This implies that the equilibrium is a stable sink. See Fig S5. %\ref{fig:phase}.

\subsection{Equilibrium sensitivity analysis}
Global sensitivity and uncertainty analysis using Latin Hypercube Sampling (LHS) of parameter space and Partial Rank Correlation Coefficients (PRCC) \cite{marino2008methodology} reveal that equilibrium distributions of diners and waiters depend significantly ($p<0.001$) on tip rates and cook pay. Equilibrium distributions of waiters also depend significantly on waiter pay. Note that the parameters that significantly influence these distributions describe the differences between restaurants and do not describe the system as a whole. See Fig S6. %\ref{fig:prcc1}. 

\section{Discussion}
\subsection{Tip abandonment threshold}
Suppose our restaurant is attempting to maximize hourly profit \eqref{profit} at equilibrium. We assume our restaurant is competing with a typical American restaurant that is not making dynamic changes to staff pay, menu prices, or tipping policies. Given the choices the other restaurant has made, our restaurant can choose base pay for cooks and waiters (within legal limits) and a gratuity policy. Both restaurants maintain identical menu prices to ensure the restaurants are true competitors; fine dining establishments do not typically compete with casual restaurants.

\begin{figure}[htb]
 \begin{center}
     \includegraphics[width=\textwidth]{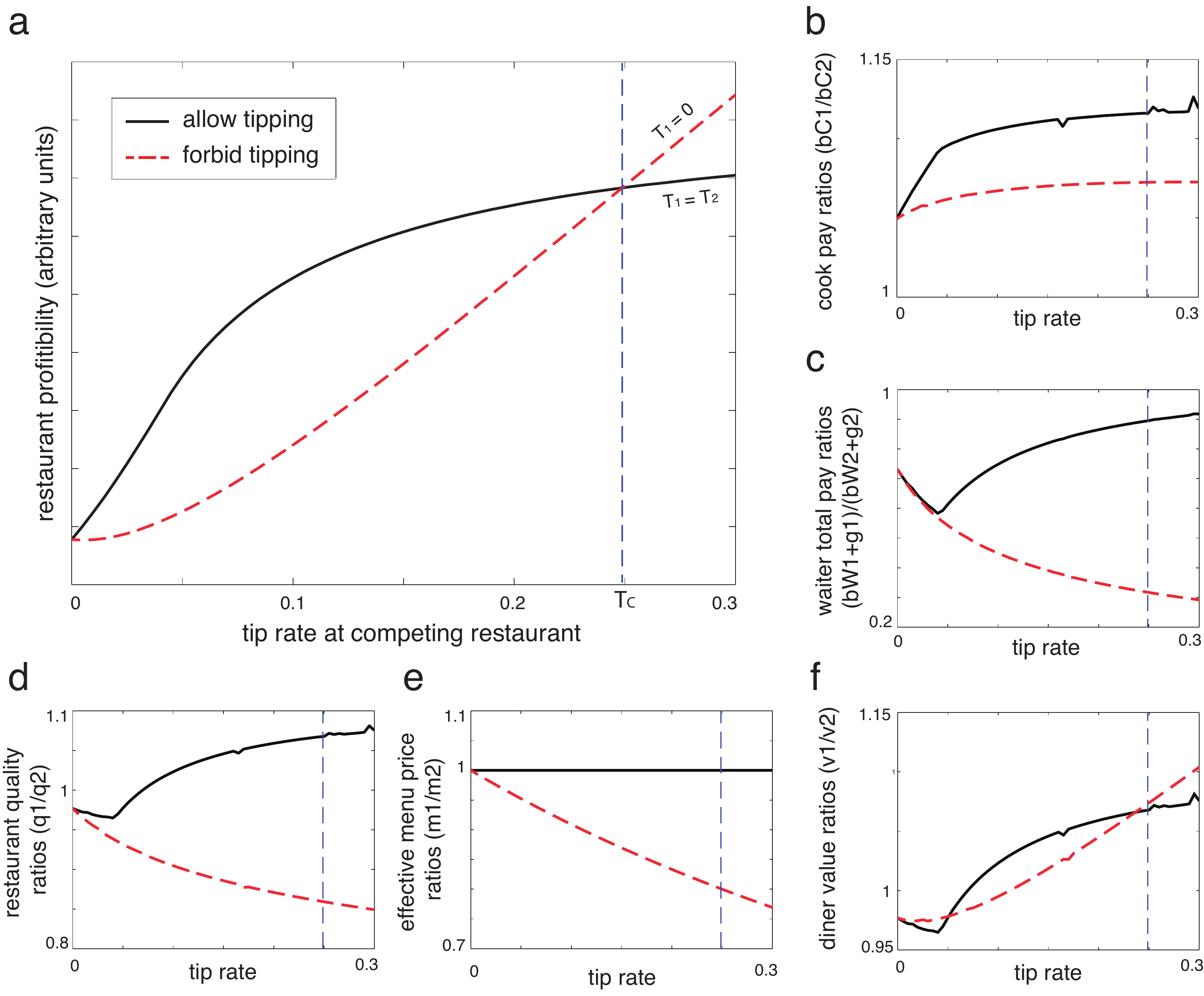} \end{center}
  \caption[Example of critical tip rate threshold $T_c$.]{(Color online) Example of critical tip rate threshold. \textbf{(a)} For conventional tip rates below some critical threshold $T_c$, a rational restaurant owner would allow diners to leave gratuity to maximize profitability (black curve). Beyond that critical threshold, a rational restaurant owner would disallow tipping in their restaurant (red dashed curve). Both curves assume that the restaurant owner selects staff pay (within legal limits) to maximize profit. For this hypothetical restaurant ecosystem, \textbf{(b)} both optimal cook pay and \textbf{(c)} total waiter pay (optimal wage plus tips) drop if we eliminate tipping. Though staff leave our restaurant in response to a no-tip policy, \textbf{(d)} the drop in perceived quality balances \textbf{(e)} the effective menu price (menu cost plus tips) near the critical tip rate. \textbf{(f)} The trade off between quality and price, as perceived by diners, drives the critical tip rate. For this example, $m_1=m_2=10, r=4, b_{W2}=10, b_{C2}=25, r_{DW}=10, r_{CW}=1$, the minimum wage for tipped workers is $2.13$, and the minimum wage for untipped workers is $7.25$.}
  \label{fig3}
\end{figure}

If the competing restaurant allows the conventional tipping rate, then there exists a critical tip rate threshold $T_c$ at which a rational restaurant owner would forbid tipping to maximize profit. Fig \ref{fig3}a shows the conventional tip rate at which a hypothetical restaurant should switch from allowing the conventional tip to abandoning tipping in their establishment. Though the critical tip rate depends on the entire restaurant ecosystem, numerical exploration indicates that the trade off between meal cost and restaurant quality, as perceived by diners, is the primary driver of the critical tip rate (see Fig \ref{fig3}b-f). Assuming the conventional tip rate continues to increase in the US, we predict that restaurants will eventually forbid tipping when it become more profitable to do so.

%%%%%%% TC FIGURE1 %%%%%%%
%%%%%%% TC FIGURE1 %%%%%%%

\begin{figure}[htb]
  \begin{center}
    \includegraphics[width=0.6\textwidth]{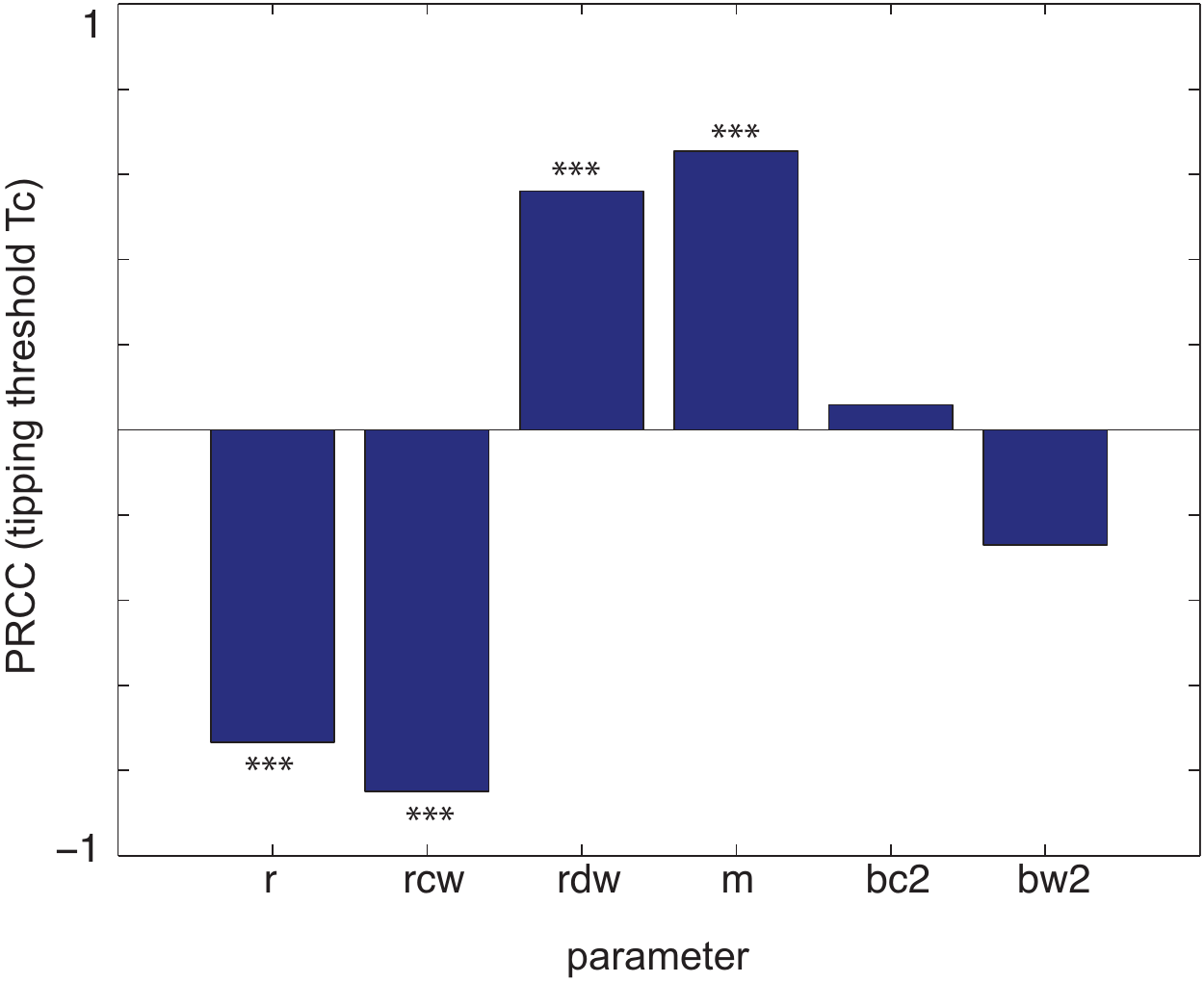} \end{center}
  \caption[Global sensitivity and uncertainty analysis for tipping threshold.]{Global sensitivity and uncertainty analysis for tipping threshold $T_c$. The Partial Rank Correlation Coefficient (PRCC) between model parameters and the tipping threshold $T_c$ are show. Asterisks indicate that the correlation is significant (***$p<0.001, N=100$ samples). Note that we use PRCC because numerical tests suggest that the relationship between parameters and $T_c$ is monotonic.}
  \label{fig4}
\end{figure}

Global sensitivity and uncertainty analysis shows that this critical tipping threshold depends significantly on the menu price shared by both restaurants, the ratio of customers to waiters and cooks to waiters, and the ratio of food quality to service quality in the eyes of the customer (see Fig \ref{fig4}). Note that the parameters that significantly influence the critical tip rate $T_c$ describe the type or ``class'' of restaurant system we are considering. For instance, fine dining restaurants maintain a low diner to waiter ratio $r_{DW}$ and high menu prices $m$. It is also likely that diners at fine dining establishments place more value on service than at casual restaurants, decreasing $r$. 

%%%%%%% PRCC FIGURE2 %%%%%%%
%%%%%%% PRCC FIGURE2 %%%%%%%

\begin{figure}[htb]
  \begin{center}
    \includegraphics[width=0.75\textwidth]{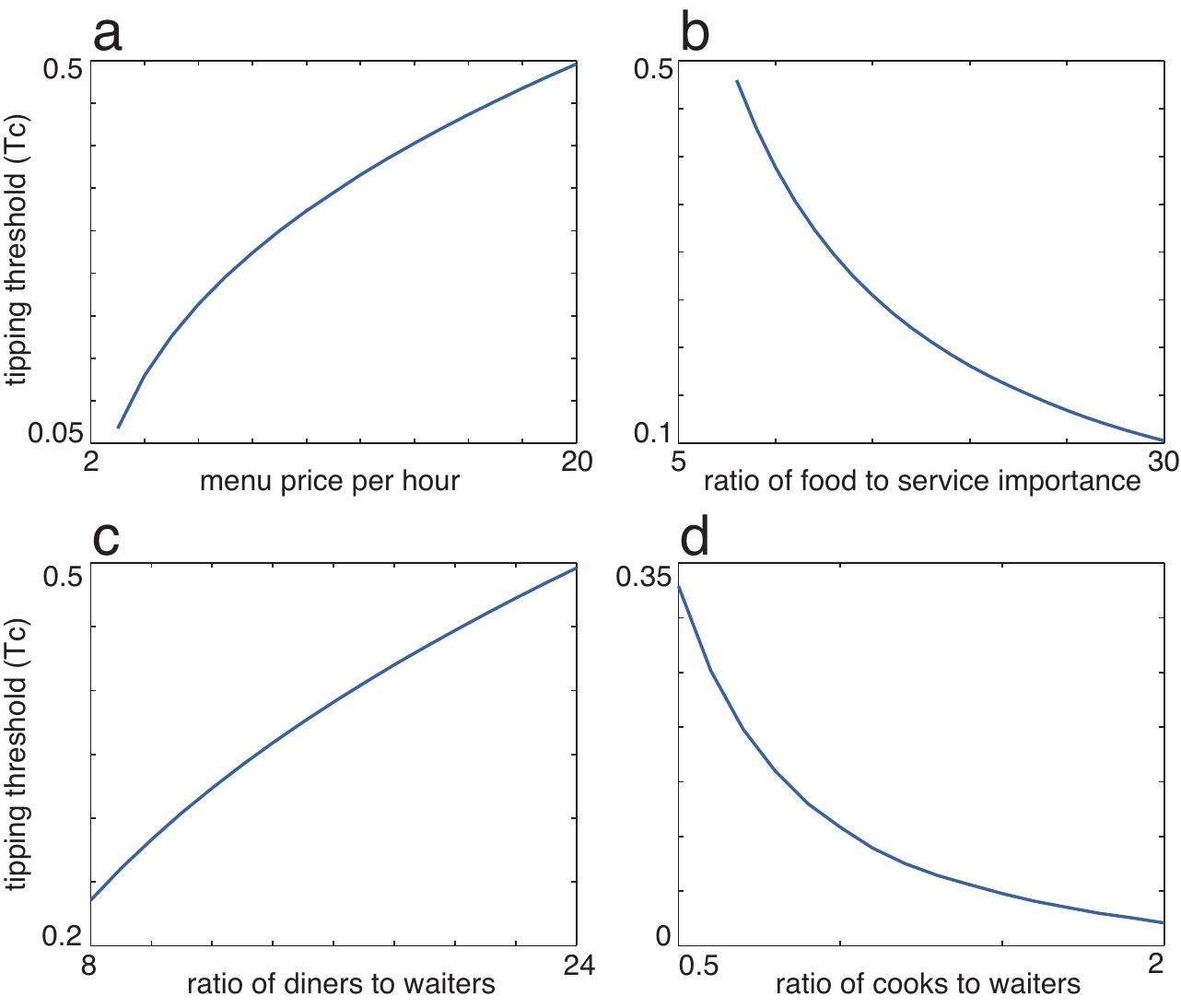} \end{center}
  \caption[Local sensitivity analysis for tipping threshold.]{Local sensitivity analysis for tipping threshold $T_c$. \textbf{(a)} Holding all else constant, higher menu price implies a larger tip threshold. This indicates that fine dining restaurants should be the last the abandon tipping if all other parameters are the same. \textbf{(b)} As diners place more relative importance on food than service, the critical tip rate decreases. Because customers at fine dining establishments likely place more value on service, we again expect that fine dining restaurants will be the last to abandon tipping. \textbf{(c)} In contrast, as the ratio of diners to waiters increases, the critical tipping rate increases. All else held constant, this would imply that casual dining establishments would drop tipping last. \textbf{(d)} As the ratio of cooks to waiters increases, the critical tip rate decreases. Though it is difficult to know this ratio for full service restaurants, it makes sense that the large cook-to-waiter ratio seen at counter service restaurants implies little to no tipping. For this example, $m_1=m_2=10, r=12, b_{W2}=5, b_{C2}=10, r_{DW}=12, r_{CW}=0.5$, unless otherwise noted on the independent axis.}
  \label{fig5}
\end{figure}

Local sensitivity analysis about `typical' American restaurant parameters suggests that increased menu price, increased service importance, increased diner-to-waiter ratio, and increased waiter-to-cook ratio all increase the critical tipping rate (see Fig \ref{fig5}). Because no type or class of restaurant increases all these parameters, we cannot say with certainty that a certain type of restaurant should abandon tipping before another. However, the three strongest correlated parameters ($r, r_{CW}, m$) support the prediction that casual dining restaurants should be the first to abandon tipping, and fine dining establishments should be the last to abandon tipping. 

%%%%%%% TC VERSUS PARAMS FIGURE %%%%%%%
%%%%%%% TC VERSUS PARAMS FIGURE %%%%%%%

This prediction is consistent with tipping practices in the most casual restaurants: customers in fast food restaurants and counter-service establishments are not typically expected to leave tips. Among restaurants that expect patrons to leave tips, the prediction is surprising because the most vocal advocates for eliminating tipping in America have been owners of upscale restaurants. However, fine dining restaurant owners cite social justice as the primary motive for eliminating tipping in their establishments \cite{estreicher2016case}. This claim is consistent with our prediction because many restauranteurs have been forced to reinstate tipping in their restaurants in order to remain profitable \cite{atlantic:notip}.

\subsection{Limitations}
As a conceptual model, system \eqref{eq:diner}-\eqref{eq:cook} cannot offer quantitative predictions with confidence. One limitation of this model is the lack of competition among many restaurants or eating at home, though this could be addressed by considering the ``competing restaurant'' as a pool of competition. Additionally, the model assumes that the benefit of more employees does not have diminishing returns. More realistically, restaurant food or service will only benefit from more employees up to a certain point; after the restaurant is fully staffed, more employees will be a waste of money and may even impede service. 

Our model also ignores both the federal law that requires restaurant owners to supplement tipped worker wages if their hourly tips do not exceed the federal minimum wage \cite{minwage2} and many state laws that impose larger minimum wages for tipped employees \cite{minwage}. We also do not provide a mechanism by which the conventional tip rate increases and merely assume that the increasing trend will continue; however, the increasing trend is supported by theoretical economic models \cite{azar2004sustains}. Finally and most importantly, this model assumes that humans behave rationally when spending or earning money, a false assumption common among economic models \cite{mcfadden2013new}. Restaurant owners may choose to abandon or maintain tipping regardless of profit, citing economically irrational reasons or responding to irrational customer feelings.

In spite of these limitations, the qualitative prediction that a critical tipping threshold exists at which restaurant owners may abandon tipping is supported by previous trends both in America and internationally. Tipping has gone in and out of fashion around the world, and though customers normally drive the introduction (or reintroduction) of the trend, restaurant owners or governments typically end the practice \cite{lynn2006tipping}.

%NEXT STEPS/PROBLEMS
\section{Conclusion}
The conceptual model presented here takes a new direction towards understanding the complex service industry. The oscillating popularity of tipping has previously been attributed to social contagion and irrational responses to classism. Using a new approach to modeling the social convention of tipping, we show that rational decisions to maximize profit may drive the cycle of the tipping trend. We predict that there exists a critical tip rate threshold at which restaurant owners would be wise to eliminate tipping in their establishments. Furthermore, we expect that casual restaurants should be the first to abandon tipping, and fine dining restaurants should be the last.

The simplicity of the model does not allow for quantitative predictions, such as when tipping will go out of fashion in the US or what the threshold tip rate will be. However, the model serves as a base for more sophisticated models and could direct economic data collection to better answer quantitative questions. This effort would be important not only to restaurant owners, but also to economists, sociologists, policy makers, and all people who play a role in or interact with the service industry.

\section{Supplementary Material}
See supplementary material for additional discussion and figures.

\section{Acknowledgments}
The authors thank Michael Lynn for sharing both data and ideas. This work was funded in part by the National Science Foundation Graduate Research Fellowship No.~1324585. Funding was also provided by the Northwestern University Undergraduate Research Assistant Program. The funders had no role in study design, data collection and analysis, decision to publish, or preparation of the manuscript.

\section{Data availability}  
All software (Matlab .m files) used to simulate the model are publicly available from the Northwestern ARCH repository (DOI:10.21985/N2DM26) at \url{https://arch.library.northwestern.edu/concern/generic_works/gf06g273j}.

\bibliographystyle{ieeetr}
\bibliography{tipLibrary_v2}

\newpage
\setcounter{figure}{0}
\renewcommand{\thefigure}{S\arabic{figure}}

\setcounter{section}{0}
\renewcommand{\thesection}{S\arabic{section}}

\section{Optimal waiter base pay}
\begin{figure}[htb]
  \begin{center}
    \includegraphics[width=0.65\textwidth]{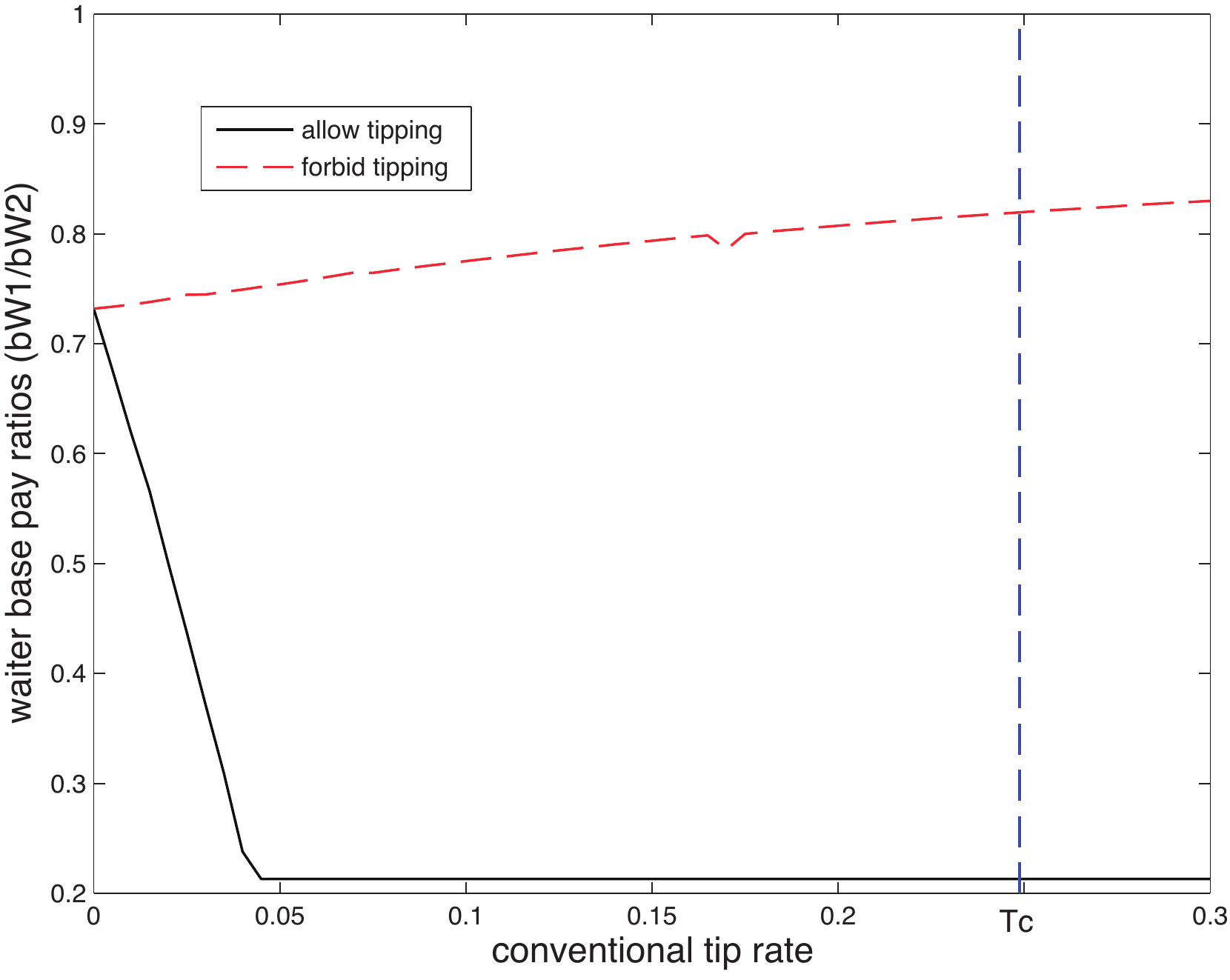} \end{center}
  \label{fig:waiterbasepay}
  \caption{Illustration of optimal waiter base pay under different tipping protocols. When our restaurant chooses to forbid tipping, the optimal waiter base pay rate increases steadily with the conventional tip rate. In other words, restaurant owners must compensate for lost tips. For this example, $m_1=m_2=10, r=4, b_{W2}=10, b_{C2}=25, r_{DW}=10, r_{CW}=1$, the minimum wage for tipped workers is $2.13$, and the minimum wage for untipped workers is $7.25$. Note that the base pay ratio is low when tipping is allowed because waiters at the other restaurant are paid $\$10$ an hour no matter what, and the optimization algorithm tells us to pay waiters minimum wage ($\$ 2.13$).}
\end{figure}

The profit optimization algorithm selects the most profitable staff pay rate for every tip rate scenario, so restaurant owners that abandon tipping must pay waiters a higher wage to compensate for the lost tips. Fig \ref{fig:waiterbasepay} shows that the relative optimal base pay when we forbid tipping must steadily increase as the tip rate increases. 

%%%%%%% FIGURE %%%%%%%
%%%%%%% FIGURE %%%%%%%

\begin{figure}[htb]
  \begin{center}
    \includegraphics[width=0.65\textwidth]{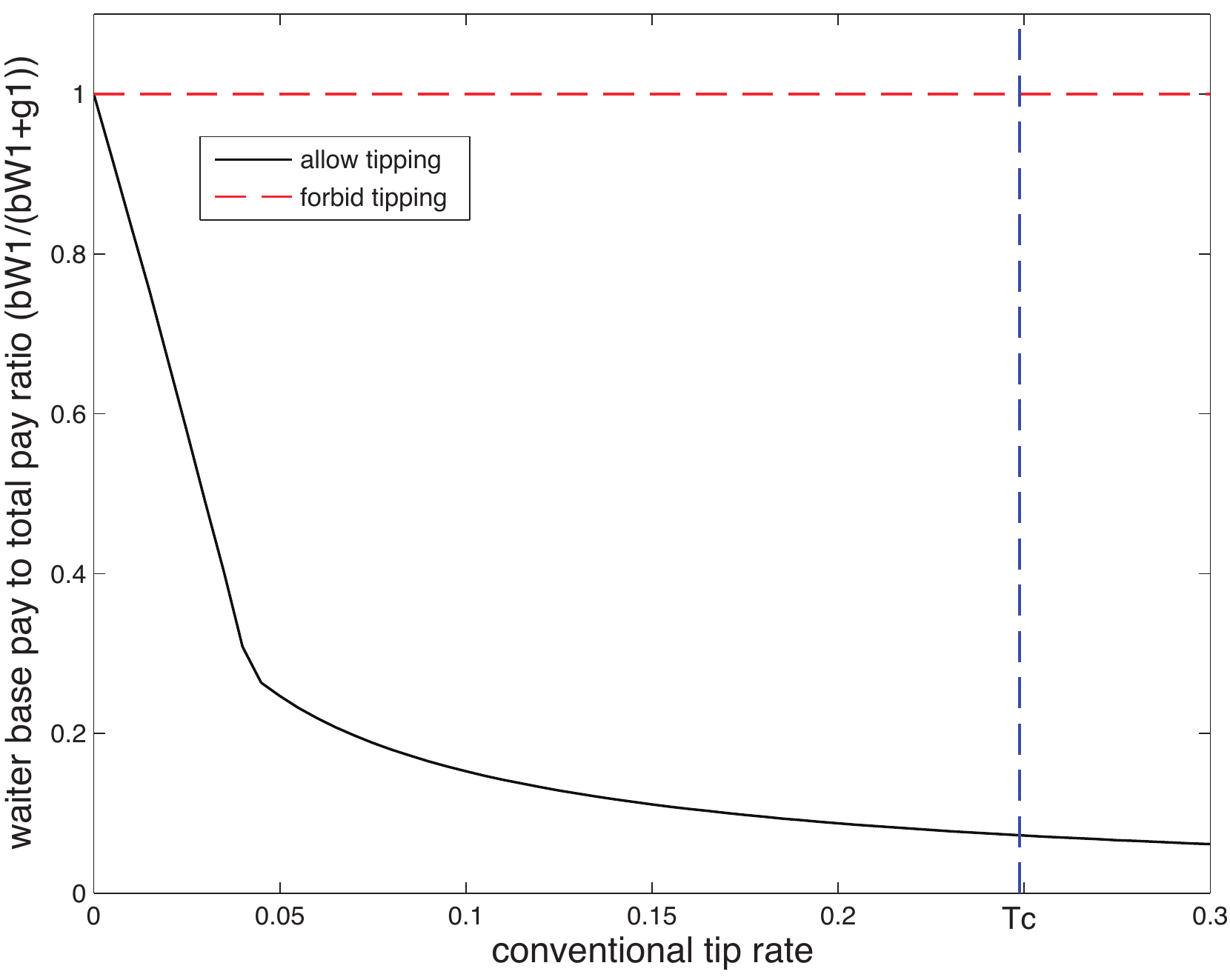} \end{center}
  \label{fig:waitertotalpay}
  \caption{Illustration of base pay as a fraction of total pay under different tipping protocols. When our restaurant allows tipping, less and less waiter pay comes from wages as the conventional tip rate increases. For this example, $m_1=m_2=10, r=4, b_{W2}=10, b_{C2}=25, r_{DW}=10, r_{CW}=1$, the minimum wage for tipped workers is $2.13$, and the minimum wage for untipped workers is $7.25$.}
\end{figure}

As the tip rate increases, waiters receive relatively more compensation from tips than from base pay (see Fig \ref{fig:waitertotalpay}). This is expected, though it is difficult to know if our ratios are realistic because tips, often paid in cash, may be under-reported. 

%%%%%%% FIGURE %%%%%%%
%%%%%%% FIGURE %%%%%%%

\section{Alternative interpretations of restaurant quality}
Though we have framed the model as counting the literal number of waiters and cooks in the restaurant, the model could be reinterpreted as measuring the total waiter and cook quality. Because the number of cooks and waiters in our restaurant is actually the number who wish to work in our restaurant, one could argue that $W$ and $C$ are indicators of the quality of staff we'll have when we hire the appropriate number of waiters and cooks.

Without any adjustment to the model, $C$ could be interpreted as the fraction of cooks who wish to work in our restaurant, and $b_{C1} C$ (total pay for all cooks) could be the quality of the cooks actually hired. We simply assume that we hire the best cooks from the pool of applicants and pay them appropriately. For instance, if ten cooks want to work at our restaurant, then we could hire all ten at $\$10$ an hour. Equivalently, we could hire only the most qualified five cooks and pay them $\$20$ an hour.

We can apply the same logic to waiters, but the model will need to be adjusted slightly. As is, gratuities are split among all waiters, and a new model would need to account for the actual number of waiters working in the restaurant. For instance, if we hire half as many waiters that are twice as talented and pay them twice the salary, our profit function would not change. However, each waiter would receive double the tips because the gratuities would be split among half as many waiters.

\section{Alternative formulations for restaurant quality}
\begin{figure}[htb]
  \begin{center}
    \includegraphics[width=0.65\textwidth]{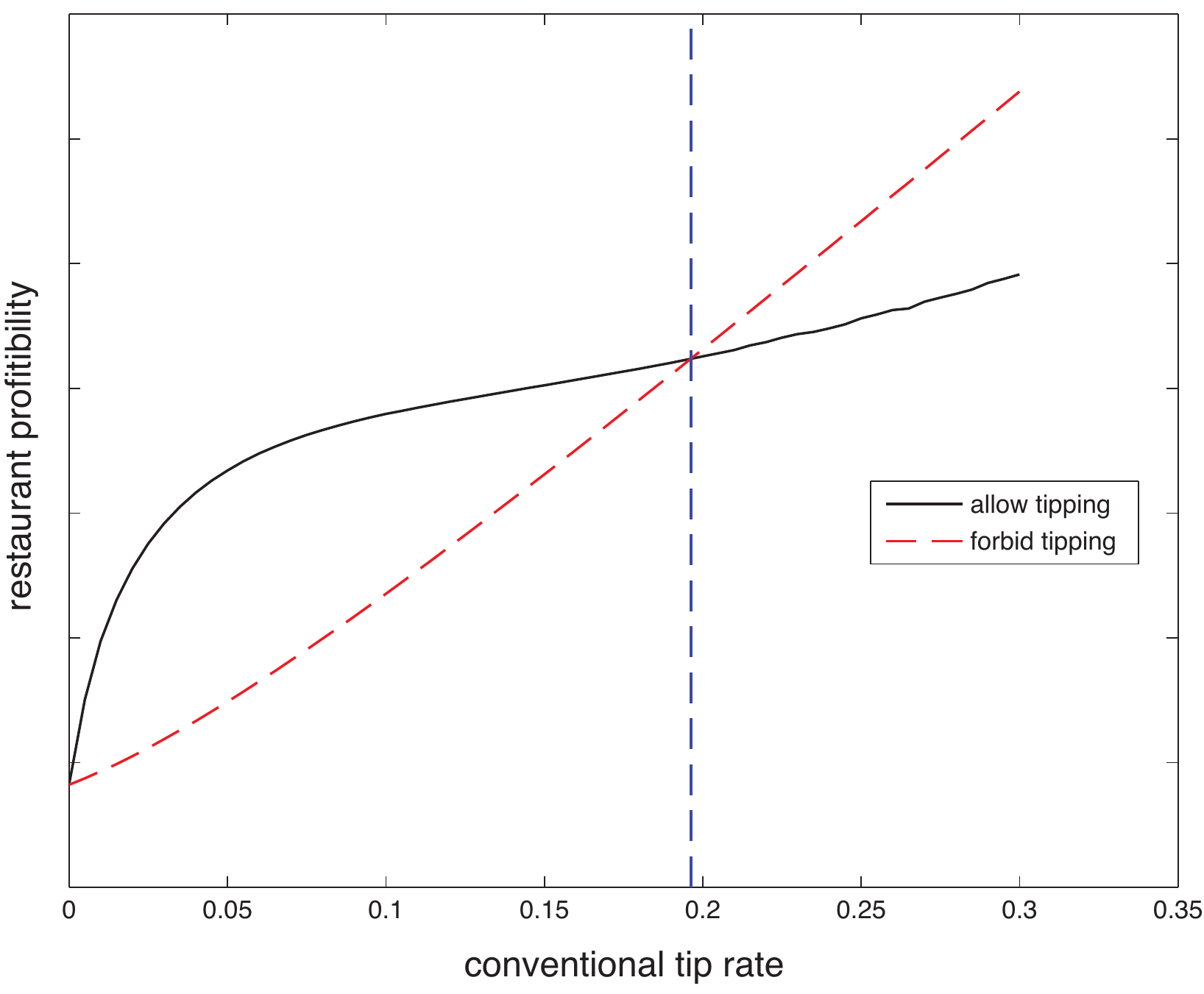} \end{center}
  \caption{Example of critical tip rate threshold with pay-dependent quality measure \eqref{eq:qalt1}. For conventional tip rates below some critical threshold (blue dashed line), a rational restaurant owner would allow diners to leave gratuity to maximize profitability (black curve). Beyond that critical threshold, a rational restaurant owner would disallow tipping in their restaurant (red dashed curve). Both curves assume that the restaurant owner selects staff pay (within legal limits) to maximize profit. For this example, $m_1=m_2=10, r=2, b_{W2}=10, b_{C2}=25, r_{DW}=10, r_{CW}=1$, the minimum wage for tipped workers is $2.13$, and the minimum wage for untipped workers is $7.25$.}\label{fig:alt1}
\end{figure}

The most difficult modeling task for this restaurant ecosystem is formulating a simple, but realistic, restaurant quality measure. The original model uses the weighted sum of waiters and cooks at our restaurant as a proxy for restaurant quality. Alternatively, a weighted sum of waiter and cook pay could be a proxy for restaurant quality:
\begin{align} \label{eq:qalt1}
q_1 = \alpha_W (b_{W1} + g_1) + \alpha_C b_{C1}
\end{align}
With this new definition of restaurant quality, the critical tip threshold remains (see Fig \ref{fig:alt1}).

%%%%%%% FIGURE %%%%%%%

\begin{figure}[htb]
  \begin{center}
    \includegraphics[width=0.65\textwidth]{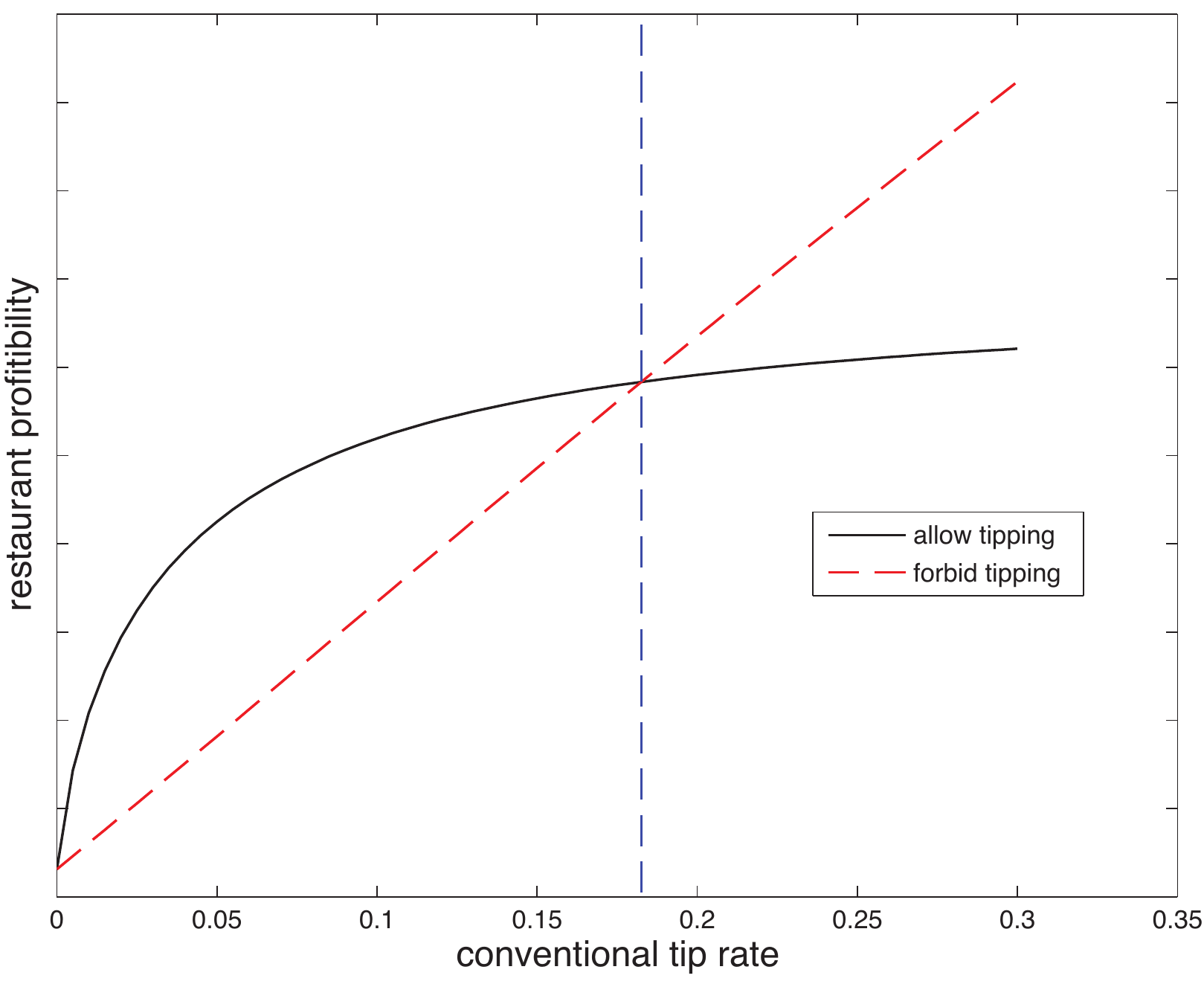} \end{center}
  \caption{Example of critical tip rate threshold with pay- and staff-dependent quality measure \eqref{eq:qalt2}. For conventional tip rates below some critical threshold (blue dashed line), a rational restaurant owner would allow diners to leave gratuity to maximize profitability (black curve). Beyond that critical threshold, a rational restaurant owner would disallow tipping in their restaurant (red dashed curve). Both curves assume that the restaurant owner selects staff pay (within legal limits) to maximize profit. For this example, $m_1=m_2=10, r=4, b_{W2}=10, b_{C2}=25, r_{DW}=10, r_{CW}=1$, the minimum wage for tipped workers is $2.13$, and the minimum wage for untipped workers is $7.25$.}\label{fig:alt2}
\end{figure}

Another alternative proxy for restaurant quality could depend on the product of staff pay and number of staff:
\begin{align} \label{eq:qalt2}
q_1 = \alpha_W \tilde{W} (b_{W1} + g_1) + \alpha_C \tilde{C} b_{C1}
\end{align}
With this new definition of restaurant quality, the critical tip threshold again remains (see Fig \ref{fig:alt2}). This suggests that the qualitative prediction that a critical tip rate exists is robust to the exact model formulation for restaurant quality.

%%%%%%% FIGURE %%%%%%%

%\section{Staffing sweet spot}
%Because we assume that restaurant quality is approximated by the linear combination of the number of waiters and cooks that wish to work at our restaurant, we cannot allow staff fractionations to differ too much from a 50/50 split (the staffing ``sweet spot"). If the vast majority of cooks or waiters work at one restaurant, then that restaurant will be grossly over-staffed and the other will be grossly under-staffed. In this case, neither restaurant will function properly, and the restaurant quality approximation will be poor. Within our parameter bounds in Table I of the main paper, we see fractionations for diners and staff as great as approximately 70/30. 

\clearpage
\newpage
\section{Additional figures}

%%%%%%% FIGURE %%%%%%%
\begin{figure}[htb]
  \begin{center}
    \includegraphics[width=0.6\textwidth]{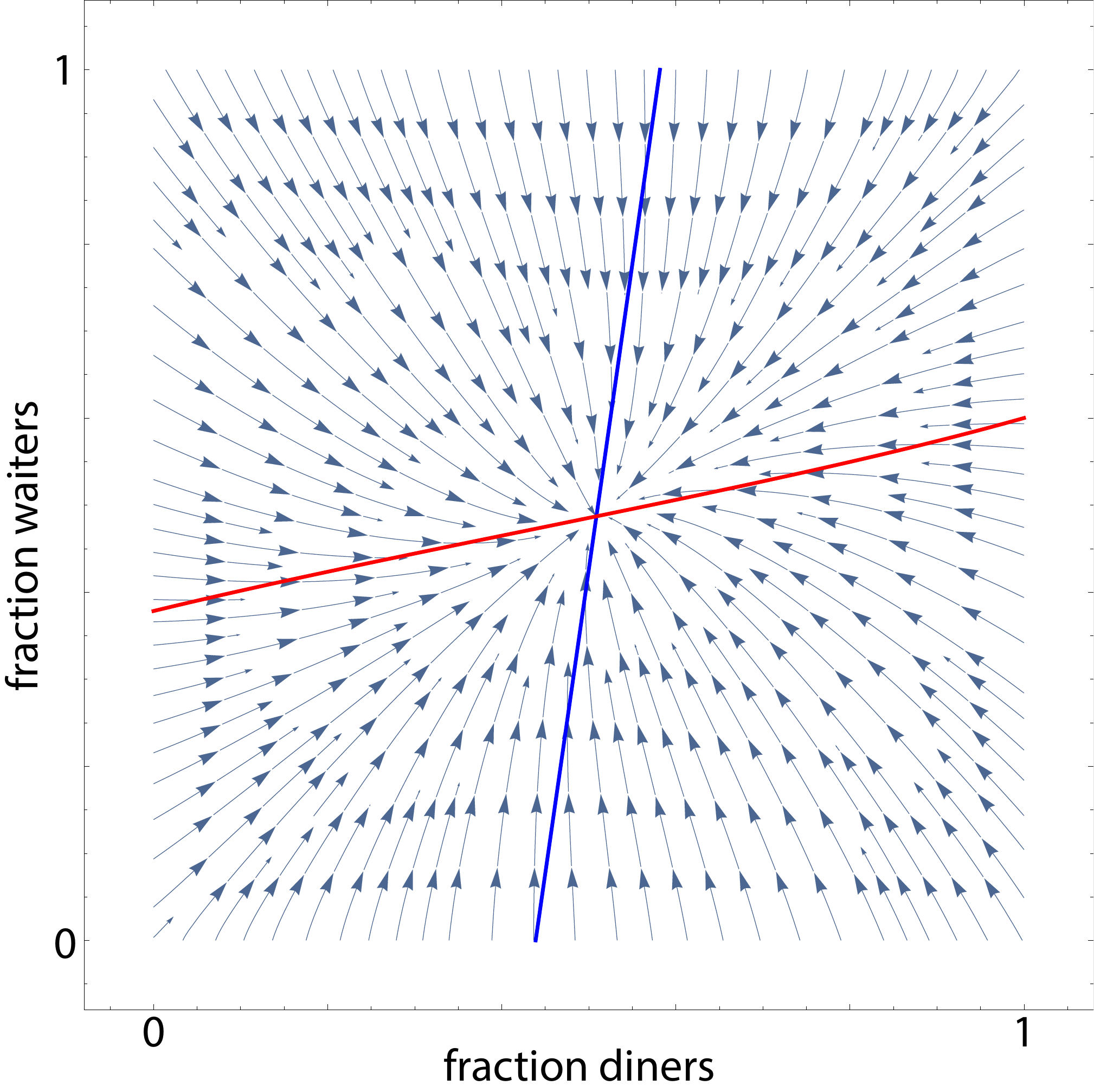} \end{center}
  \label{fig:phase}
  \caption{Phase portrait of two identical restaurants with differing tip rates. Our restaurant (shown) enforces an automatic gratuity of $T_1=0.15$, and the competing restaurant allows the conventional tip rate of $T_2 = 0.2$. The steady state is $(D^*,W^*,C^*) = (0.51,0.49,0.50)$. Nullclines $\mathrm{d} D/ \mathrm{d} t = 0$ (blue) and $\mathrm{d} W/ \mathrm{d} t = 0$ (red) are superimposed. For this example, $m_1=m_2=10, r=12, b_{W1}=b_{W2}=5, b_{C1}=b_{C2}=10, r_{DW}=1, r_{CW}=1$.}
\end{figure}
%%%%%%% STREAMPLOT FIGURE %%%%%%%

%%%%%%% PRCC FIGURE1 %%%%%%%
\begin{figure}[htb]
\begin{center}
\includegraphics[width=0.9\textwidth]{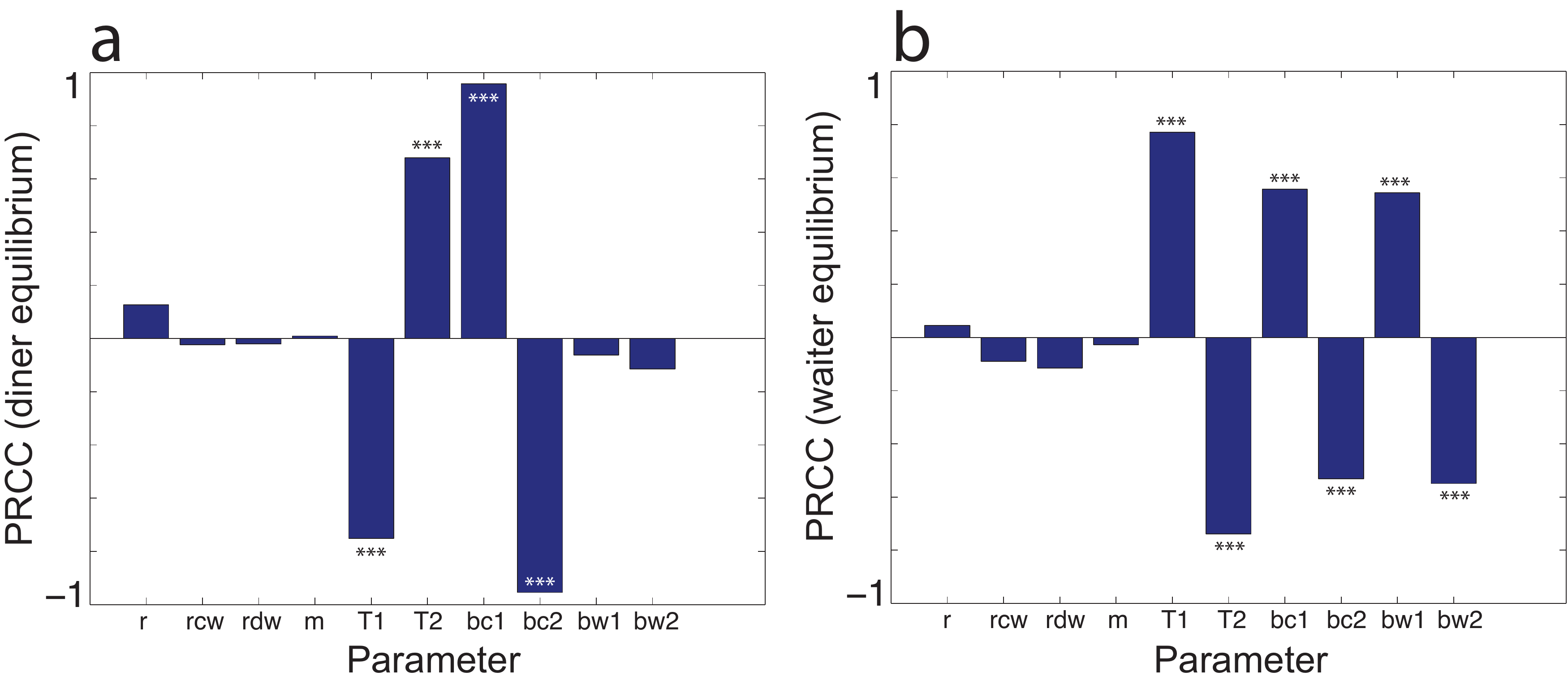} \end{center}
   \label{fig:prcc1}
  \caption{Global sensitivity and uncertainty analysis for equilibrium state. \textbf{(a)} Partial Rank Correlation Coefficient (PRCC) between model parameters and diner equilibrium. \textbf{(b)} PRCC between model parameters and waiter equilibrium. Asterisks indicate that the correlation is significant (***$p<0.001, N=100$ samples). Note that we use PRCC because numerical tests suggest that the relationships between parameters and equilibria are monotonic.}
\end{figure}
%%%%%%% PRCC FIGURE1 %%%%%%%

\end{document}